\documentclass[aps,prb,showpacs,twocolumn,floats]{revtex4}
\usepackage{amssymb}
\usepackage{amsbsy}
\usepackage{amsmath}
\usepackage{graphicx}
\usepackage{amsmath}
\usepackage{times}
\usepackage{color}
\usepackage{setspace}
\usepackage{bm}% bold math
\newcommand\bea{\begin{eqnarray}}
\newcommand\eea{\end{eqnarray}}
\newcommand\beq{\begin{equation}}
\newcommand\eeq{\end{equation}}

\begin{document}
%\hskip 2 in\txtbf{\red{Highlighted}}

\title{Andreev tunneling and Josephson current in light irradiated graphene}

\author{Debabrata Sinha and Satyaki Kar\footnote{Corresponding author:satyaki.phys@gmail.com}}

\affiliation{Theoretical Physics Department, Indian Association for
the Cultivation of Science, Jadavpur, Kolkata-700032, India.}

\date{\today}

\begin{abstract}
We investigate the Andreev tunneling and Josephson current in graphene irradiated with high-frequency linearly polarized light. 
The corresponding stroboscopic dynamics can be solved using Floquet mechanism which results in an effective stationary theory to 
the problem exhibiting an anisotropic Dirac spectrum and modified pseudospin-momentum locking. 
When applied to an irradiated normal graphene - superconductor (NS) interface, such analysis reveal Andreev reflection (AR) to become an oscillatory 
function of the optical strength. Specifically we find that, by varying the polarization direction we can both suppress AR 
considerably or cause the Andreev transport to remain maximum at sub-gap excitation energies even in the presence of Fermi level 
mismatch. Furthermore, we study the optical effect on the Andreev bound states (ABS) within a short normal-graphene sheet, sandwiched between two $s$-wave 
superconductors. It shows redistribution of the low energy regime in the ABS spectrum, which in turn, has major effect in shaping the Josephson 
super-current. Subjected to efficient tuning, such current can be sufficiently altered even at the charge neutrality point. 
Our observations provide useful feedback in regulating the quantum transport in Dirac-like systems, achieved
via controlled off-resonant optical irradiation on them.
\end{abstract}
\maketitle
\section{Introduction}
\label{intro}

The quantum transport in graphene\cite{Beenakker-PRL06,ben-rmp}, with its low energy Dirac 
spectrum\cite{Beenakker-PRL06,ben-rmp,neto,sdsharma} at the edges of the brillouin zone, has remained an engaging field of study ever since its inception 
in 2004\cite{andre}. Though experimental difficulties still remain in detecting its transport
characteristics at the edges\cite{bastos}, its bulk behavior,
described by the massless, chiral fermions\cite{sdsharma} has been quite well probed by now.
Tuning such system with straining or introducing gap in the Dirac spectrum
witness noticeable variations in the charge transport. Particularly, graphene based superconductor|normal|superconductor (SNS)
junction can be tuned to enhance supercurrent at the charge neutrality point upon straining\cite{lind2} or an energy gap can enhance the 
pseudospin inverted Andreev conductance in a graphene-based superconductor/pseudoferromagnet junction\cite{majidi}.
As it turns out, such modifications can be easily implemented via optical irradiation.
The transport properties of the Dirac-like systems are very much susceptible to light irradiation and produce
interesting outcomes such 
as exciting surface plasmon polaritons\cite{farhat,plasmon,plasmon2} in graphene sheet, enhancing controllability of 
electrodynamics in graphene-based metamaterials\cite{iorsh,kuzmin} or allowing photo-reduction of graphene-oxide films to 
tune wettabiity\cite{furio} and so on. 

The energetics of the charge carriers of a graphene monolayer, periodically driven via high frequency 
electromagnetic light waves with electrons strongly coupled to the photons, has been analyzed recently\cite{Shelykh-1,Shelykh-2,
Shelykh-3,Shelykh-4,Shelykh-5,nori,zhou} using Floquet theory. There, the electrons get 
dressed by the field exhibiting drastic changes in the dispersion. For example, it is observed that the 
circularly polarized light create a field induced gap at the charge neutrality or Dirac point \cite{Luis-1, Luis-2, Jin-2}. In 
contrast the energy spectrum of electrons dressed by linearly polarized light is modified by Bessel function and it 
remains gapless. The physical properties of dressed electrons have been studied in various condensed matter 
systems including quantum well \cite{Kibis-1,Kibis-2,Kibis-3}, quantum rings \cite{Kibis-4} and in recently in Dirac 
materials like graphene \cite{Shelykh-2,Jin-2,Jin-1}, Weyl semimetals \cite{Rubio} and topological 
insulators\cite{sinha}. In graphene related systems, particularly attentions paid to the transport properties
of dressed electrons in $p-n$ junctions \cite{Jin-1}, magneto-transport \cite{Shelykh-1} and spin transport of 
dressed electrons\cite{Shelykh-3}, optical response of dressed electrons \cite{Shelykh-5} and field induced
topological phase transition\cite{Jin-1, Ezawa, oka}.

Though an effective stationary theory is constructed for stroboscopic evolution of the fermionic 
wavefunction under light irradiation, to the best of our knowledge, no study on transport behavior of the light 
irradiated superconducting graphene junctions has been performed yet. We
would like to bridge this gap in the literature and study the tunneling conductance of a normal metal-superconductor 
(NS) junction as well as the Andreev bound state (ABS) and Josephson current of a SNS junction in a strongly irradiated
graphene sheet. As a primary investigation, this paper deals with only irradiation via linearly polarized light. 
A graphene is a semimetal in which superconductivity can be induced via proximity effect \cite{proxi}. Transport through a NS junction 
experiences Andreev reflection (AR) for energy-bias smaller than the superconducting gap\cite{linder}. The
resulting electron-hole conversion in the Normal (N) sub-system and the cooper pair production in the superconductor (S) counterpart 
develops a finite conductance across the NS junction. An irradiation via linearly polarized light offers a tuning parameter 
$\alpha$ (which is a function of both intensity and frequency of the light, to be elaborated later on) to the problem. 
In the off-resonant conditions, the quasiparticle velocities along a direction normal to the irradiated field 
get reduced, from its original graphene Fermi velocity $v_F$, by a factor of Bessel's function $J_0(\alpha)$. 
Thus the low energy spectrum of graphene becomes anisotropic, tunable by the optical parameter
$\alpha$. This tunability allows for considerable variation in the Andreev current and subgap conductance becomes an oscillatory 
function in $\alpha$. With rotation of the plane of polarization, this current can be enhanced to maximum or
suppressed considerably, as can be found from our calculation and results in section III.

In this work we also study the Andreev bound states (ABS) and Josephson current on a light irradiated SNS 
junction in graphene. The analysis of ABS and Josephson current in graphene SNS junction is well studied in the literature\cite{titov,maiti,jacob}. 
Our objective is to probe the effect of linearly polarized light on such system/assembly. 
We find that the low energy spectrum of the ABS get sufficiently affected by the optical irradiation.
As a result, the Josephson current get enhanced or suppressed depending on the direction of polarization as well as the value
of the chemical potential.
Signature of such modifications are found at the charge neutrality point as well, even though the density of state vanishes there.
These interesting observations, in fact, can provide possible route to control
quantum transport in graphene with relevance to the spintronic based applications.

We reiterate here that the Hamiltonian of our light irradiated system is time periodic due to the presence of a
time dependent field of polarization and
 we resort to the Floquet formalism to analyze the stroboscopic dynamics\cite{nori,nori2} of this time-periodic problem.
Recently several authors have used the Floquet theory in the context of Dirac materials\cite{Shelykh-1,Shelykh-2,Shelykh-3,sinha,oka}. It generally presumes the frequency of the optical 
field to be off-resonant and thus does not cause any direct electronic transition\cite{Jin-1}. This can be achieved if the photon 
energy of the polarized light meets the condition $\omega\tau_0>>1$ where $\tau_0$ is the relaxation time of the unirradiated 
graphene\cite{Shelykh-2}. In the off-resonant condition,
energy conservation can thus be respected within a first order perturbation theory resulting in an effective stationary Hamiltonian
of the problem. Our construction, in presence of a linearly polarize light, closely follows the work presented in Ref.\onlinecite{Shelykh-3}.

The rest of the paper is organized as follows. In section II, we describe the Floquet theory to derive the stationary Hamiltonian that we later work on.
In section III we describe the Andreev transport in irradiated graphene NS junction. In section IV, we describe ABS and Josephson current 
through the corresponding SNS junction and finally in section V, we summarize our work and conclude.

%We should mention here that a time dependent field, as in our case, takes the system out of equilibrium and we confine ourselves only in the stroboscopic dynamics\cite{kris} where a 
%Floquet theory\cite{debu} helps us getting an effective stationary Hamiltonian for the problem. Thus seeing stroboscopically, we deal with an effetive static system whose observables can be obtained in conventional ways.

%The paper is organized as follows. In section I, we describe the Floquet theory to derive the stationary Hamiltonian that we later work on.
%In section III, we describe the transport through a NS junction and in section III, we describe the SNS junction and the Andreev bound states
%and Josephson currents that contributes to the transport. Finally in section IV, we summarize our work and conclude.

\section{Effective stationary theory}

As mentioned in the Introduction, what follows below for the derivation of the stationary Hamiltonian of our irradiated problem
is an simple extension of the work performed in Ref.\onlinecite{Shelykh-1,Shelykh-2,Shelykh-3,Shelykh-4}.

The low energy physics in graphene, around the Dirac point, are described by the linear Hamiltonian 
$H_{\bf k}=\hbar v_F{\bf \sigma}.{\bf k}$ where ${\bf\sigma~\&~k}$ denote the pseudo-spin (originating from the two sublattice indices in the underlying 
honeycomb lattice)  and the wave-vectors of the Dirac particles respectively.
 {In presence of a polarizing field, the canonical momentum gets the Pierel's substitution yielding $H_{\bf k}= \hbar v_F{\bf \sigma.(k+eA)}$,
where ${\bf A}$ denotes the magnetic vector potential.
}
{Electrons/holes get dressed by the field\cite{Shelykh-1} and those quasiparticles,
% And thus in this picture, the field B component need not to be considered separately, 
% though photovoltaic Hall effect can be seen in those situations\cite{oka}.
% We should also mention here that a stationary magnetic field 
% creates discrete landau levels in the graphene Dirac spectrum and the distance between them changes with the dressing EM field\cite{Shelykh-1}. 
%However may cause the surface states to survive, depending on the type of edges\cite{dahal-10}.
}
for an electric field $E=E_0 sin(\omega t)[cos~\theta_0~{\hat x}+sin~\theta_0~{\hat y}]$,
% and the corresponding vector potential 
% $A=\frac{E_0}{\omega}cos(\omega t)(cos~\theta_0,sin~\theta_0)$ in the $xy$ plane 
are described by the Hamiltonian
\begin{eqnarray}
H_{\bf k} &=& \hbar v_F[\sigma_xk_x+\sigma_yk_y]+\nonumber\\
&&\frac{ev_FE_0cos(\omega t)}{\omega}[cos~\theta_0~\sigma_x+sin~\theta_0~\sigma_y],
\label{fermhamden}
\end{eqnarray}
%(as $k_z=0$)
 with the Schr\"{o}dinger equation given by, $i{\dot \psi}_{\bf k}=H_{\bf k}\psi_{\bf k}$.

In the basis of spinor $s_+=(1~~0)^T$ and $s_-=(0~~1)^T$, we have $\sigma_zs_\pm=\pm s_\pm$.
At the Dirac point, the wave-function $\psi_{\bf k=0}$ actually corresponds 
to the non-stationary part {$H_0=\frac{ev_FE_0cos(\omega t)}{\omega}[cos~\theta_0~\sigma_x+sin~\theta_0~\sigma_y]$, which appears due to the presence of the electromagnetic 
field. Its eigenstates,compatible with the Schrodinger equation, are given by  
\begin{eqnarray}
 \psi_0^\pm=\frac{1}{\sqrt{2}}[ e^{-i\theta_0}s_+\pm s_-]e^{\mp i(\alpha/2)sin(\omega t)},~\rm{where~\alpha=\frac{2ev_FE_0}{\hbar\omega^2}}\nonumber
\end{eqnarray}
and they represent the time-dependent basis for the problem.
So the general wave-function can be written as,
\begin{align}
 \psi_{\bf k}= a_{\bf k}^+(t)\psi_0^++a_{\bf k}^-(t)\psi_0^-~~ (=c_+(t)s_++c_-(t)s_-)
\end{align}
where the coefficients of the two basis are related as 
\begin{equation}
\begin{pmatrix}
c_+(t)\\
c_-(t)\end{pmatrix}=\begin{pmatrix}
e^{-i\theta_0}\\
1\end{pmatrix}[a_{\bf k}^+(t)e^{-i\frac{\alpha}{2}sin(\omega t)}\pm a_{\bf k}^-(t)e^{i\frac{\alpha}{2}sin(\omega t)}].\nonumber
\end{equation}
Solution to the Schrodinger equation, in the time-dependent basis then gives
$i\hbar \partial \psi_{\bf k}/\partial t=H_{\bf k}\psi_{\bf k}~~i.e.,$
\begin{eqnarray}
 i{\dot a}_{\bf k}^\pm(t)&=&\pm v_F[\{k_xa_{\bf k}^\pm(t)+ ik_ye^{\pm i\alpha sin(\omega t)}a_{\bf k}^\mp(t)\}cos~\theta_0
 \nonumber\\&&+\{k_ya_{\bf k}^\pm(t)- ik_xe^{\pm i\alpha sin(\omega t)}a_{\bf k}^\mp(t)\}sin~\theta_0].
 \label{fl1}
\end{eqnarray}
}

{Let us now bring in the Floquet picture for this periodically driven system, which says that for stroboscopic evolution we can write
$\psi_{\bf k}(t+T)=e^{-i\epsilon_{\bf k} T}\psi_{\bf k}(t)$, $T=2\pi/\omega$ being the time period of the field.
Here $\epsilon_{\bf k}$ is the quasi-energy of the Floquet mode which turns out to be the eigenvalue of the corresponding 
Floquet Hamiltonian.
We can absorb this exponential dependence in the coefficients $a_{\bf k}^\pm$ and write a frequency-Fourier transform as 
\begin{equation}
 a_{\bf k}^\pm(t)=e^{-i\epsilon_{\bf k} t}\sum_n \tilde a_{{\bf k},n}^\pm e^{in\omega t}.\nonumber
\end{equation}
With this, Eq. \ref{fl1} becomes
\begin{eqnarray}
 &&(\epsilon_{\bf k}-n\omega)\tilde a_{{\bf k},n}^\pm=\pm v_F[\{k_x\tilde a_{{\bf k},n}^\pm+ i\sum_{n'}J_{n-n'}(\pm\alpha)k_y
 \tilde a_{{\bf k},n'}^\mp\}\nonumber\\
 &&cos~\theta_0+\{k_y\tilde a_{{\bf k},n}^\pm- i\sum_{n'}J_{n-n'}(\pm\alpha)k_x
 \tilde a_{{\bf k},n'}^\mp\}sin~\theta_0]
\label{fl-eq}
 \end{eqnarray}
where we utilize Jacoby-Anger formula, $e^{ixsin(t)}=\sum_{m=-\infty}^\infty J_m(x)e^{imt}$ with $J_m(x)$ denoting the Bessel's function of first kind.
Now  we consider only the 1st Floquet zone as the Floquet replicas corresponding to $n\ne0$ can be disregarded as long as 
$\omega$ is large enough compared to the frequencies corresponding to any direct electronic transition between the conduction
electrons.
% For high frequency irradiation with $\hbar\omega\sim meV$, this is a reasonable assumption given that the kinetic energy of the 
% quasiparticles in graphene are $\hbar v_Fk\sim 10^{-7}~ eV$.
Next, it is evident that at high frequency or very small $E_0$ ($i.e.,$ small $\alpha$), $J_{n'}(\alpha)$ is dominant for $n'=0$. 
Also the quantum amplitude $\tilde a_{{\bf k},n'}^\pm$, for $n'\ne0$, correspond to emission/absorption of $n'$ photons by the 
electrons and hence smaller compared to the $a_{{\bf k},0}^\pm$. These two conditions, together, justifies the second approximation
of considering only the $n'=0$ term in Eq. \ref{fl-eq}. 
It basically relies on the limit 
\begin{eqnarray}
|J_{n'}(\alpha)a_{{\bf k},n'}^\pm/J_0(\alpha) a_{{\bf k},0}^\pm|<<1
\end{eqnarray}
for $n'\ne0$, as described in Ref.\onlinecite{Shelykh-1,Shelykh-2,Shelykh-3}.
Hence it excludes the points where $J_0(\alpha)\rightarrow 0$
and with the large off-resonant frequency considered, no n-photon absorption or emission process remains 
present\cite{Shelykh-1,Shelykh-2,Shelykh-3} within the approximation.
However, to keep our results more accountable, we consider only small values of $\alpha$ for drawing any conclusion from our work.
}

% in terms of the phase-band picture\cite{bhaskar}, which is a general 
% extention of the Floquet theory at an arbitrary time ($i.e.$, not necessarily stroboscopic\cite{nori}).
% With this, we can write the evolution of the state as $\psi_{\bf k}(t)=e^{-i\epsilon_{\bf k} t}\psi_{\bf k}(0)$.
% where $\epsilon_{\bf k}$ denotes the phase-band energy. 
% Assigning the time dependence in the coefficient as $a_{\bf k}^\pm(t)=e^{-i\epsilon t}\tilde a_{{\bf k}}^\pm$,

% Now for off-resonant $\omega$ ($i.e.,~~\omega$ far from electron resonant frequency \cite{Shelykh-3}), absorption of the incident wave by
% the electrons becomes very rare and the amplitudes $\tilde a_{{\bf k},n}$ for absorption (or emission) of n-photons becomes very small
% for $n\ne0$. So in the first order perturbation theory, contributions from terms involving $n\ne0$ (as well as $n'\ne n$ in Eq.~\ref{fl-eq})
% are disregarded\cite{Shelykh-1,Shelykh-2,Shelykh-3}.
{This leads us to the equation,
\begin{eqnarray}
 \epsilon_{\bf k} \tilde a_{{\bf k},0}^\pm&=&\pm v_F[\{k_x\tilde a_{{\bf k},0}^\pm+ iv_FJ_0(\alpha)k_y\tilde a_{{\bf k},0}^\mp\}
 cos~\theta_0\nonumber\\
 &&+\{k_y\tilde a_{{\bf k},0}^\pm- iv_FJ_0(\alpha)k_x\tilde a_{{\bf k},0}^\mp\}sin~\theta_0].
\label{fl-eq2}
\end{eqnarray}
Eq.~\ref{fl-eq2} is just like a stationary Schr\"{o}dinger equation with an effective Hamiltonian given by
\begin{eqnarray}
 H'&=&\hbar[\{\sigma_zv_Fk_x-\sigma_yv_FJ_0(\alpha)k_y\}cos~\theta_0\nonumber\\
 &&+\{\sigma_zv_Fk_y+\sigma_yv_FJ_0(\alpha)k_x\}sin~\theta_0]
 \label{st-eq}
\end{eqnarray}
which can be unitary transformed to get a conventional form,
\begin{eqnarray}
 H&=&\hbar v_F[\{\sigma_x k_x+\sigma_yJ_{0}(\alpha) k_y\}cos~\theta_0\nonumber\\
 &&+\{\sigma_x k_y-\sigma_y J_{0}(\alpha)k_x\}sin~\theta_0]
\label{irr-graphene}
\end{eqnarray}
}
So, the Hamiltonian of a graphene get modified when irradiated with a linearly polarized light. Eq.(\ref{irr-graphene}) shows that, 
the velocity vector is not parallel to the wave-vector (unless direction of propagation is along $x$ or $y$)
and thus quasiparticle trajectory deviates from that it would be
in absence of the light irradiation. Note that, this Floquet theory formalism can be extended to a circularly
polarized light as well. It is well established that, circularly polarized light introduces a gap at the Dirac point of graphene by breaking
the time reversal symmetry. However, as mentioned earlier, this paper confines only in studying the effects of a linearly polarized light. 
In the next following section, we use the Hamiltonian Eq.(\ref{irr-graphene}) for our light irradiated system and discuss what consequences 
it leads to in the context of Andreev transport, ABS and Josephson current. 
To be precise, all the results that are presented in this paper, correspond to two values of $\theta_0$, namely $\theta_0=0$ and $\theta_0=\pi/2$.

% {We should mention here that including terms $n'=1,-1$ turns this 2-level problem at each ${\bf k}$, to a 6-level problem
% which make the numerical work more rigorous without giving any further useful informations in large $\omega$ limit.}
% gives additional term of $-2v_FJ_1(\alpha)k_y
% sin(\omega t)\tilde a_{{\bf k}}^\mp$ in Eq. \ref{fl-eq2}, which becomes zero at stroboscopic times $t=m\frac{2\pi}{\omega}$, 
% with $m$ integer. Its only the 2nd order correction corresponding to $n'=2,-2$ that gives non-zero contribution of
% $2iv_FJ_2(\alpha)k_ycos(\omega t)\tilde a_{{\bf k}}^\mp$ at stroboscopic times.}

% {So though the Hamiltonian becomes non-stationary in general, at stroboscopic times, we get the stationary expression of
% \begin{equation}
%  H'=\hbar[\sigma_zv_Fk_x-\sigma_yv_F(J_0(\alpha)+2J_2(\alpha))k_y]
% \end{equation}
% in place of Eq. \ref{st-eq}.
% }

\section{Irradiated Graphene N-S Junction}

We consider a NS junction in an irradiated graphene sheet occupying the $x-y$ plane with normal region at $x<0$ and superconducting region at $x>0$. The superconductivity in graphene is induced via proximity effect when a superconducting electrode is kept close to a graphene sheet \cite{Beenakker-PRL06}. Either side
of the junction can be described by the Dirac-Bogoliubov-de Gennes (DBdG) equations \cite{Beenakker-PRL06}
\begin{eqnarray}
\begin{pmatrix}
\mathcal{H}_{\pm}-\mu+U(r) & \Delta(r)\\
\Delta^{*}(r) & \mu-U(r)-\mathcal{H}_{\pm}
\end{pmatrix}
\Psi_\pm=\epsilon\Psi_\pm
\label{DBdG}
\end{eqnarray}
Here, $\Psi_\pm=(u_\pm,v_\pm)$ is the 4-component fermionic wave function where the electron-like and hole-like spinors are given as 
$u_\pm=(\Psi_{A\pm},\Psi_{B\pm})$ and $v_\pm=(\Psi^{\star}_{A\pm},-\Psi^{\star}_{B\pm})$ respectively. 
$u_\pm$ and $v_\mp$ are time-reversal partner of each other,
as the Hamiltonian possess time-reversal symmetry. 
%\red{For the same reason spin indices also do not appear here.} 
The index $+(-)$ stands for two valley $K$ and $K^{'}$ points (that constitutes the Fermi surface in the undoped graphene).
$\mu$ denotes the Fermi energy, $A$ and $B$ indicates the two sublattices within the hexagonal lattice of graphene. 
{In graphene, electron and hole states are connected and can originate
from same branch of the electronic spectrum. Moreover, the quasiparticles require two-component wavefunction description to define relative contributions of the sublattices A and B\cite{klein}. This sublattice or equivalent pseudospin index
give the notion of chirality in the graphene transport.} 
The spin-singlet pair potential 
$\Delta(r)$ in Eq.(\ref{DBdG}) is modeled as $\Delta(r)=\Delta_{0}e^{i\phi}\Theta(x)$ where $\Delta_{0}$ and $\phi$ are the amplitude and 
phase of the induced superconducting order parameter, respectively. In superconducting region there is a gap in the energy spectrum 
$|\Delta|=\Delta_{0}$ at the Fermi energy. The potential $U(r)$ give the relative shift of Fermi energy between the normal and 
superconducting regions of graphene sheet and modelled by $U(r)=-U_{0}\Theta(x)$. As we discussed, linearly polarized light does not 
break the valley degeneracy of graphene and for calculations,
it suffices to concentrate only on a single valley. 

{Quantitative analysis of {{A}}ndreev tunneling in a graphene NS junction
has been done extensively in Ref. \onlinecite{Beenakker-PRL06,ben-rmp,subhro}.
As we find that a high frequency irradiation modifies only 
one component of the quasiparticle velocity, we briefly touch upon those derivations
for anisotropic {{c}}arrier velocities, in the following.}

The energy spectrum in N or S region can be 
written as
\begin{eqnarray}
\epsilon=\sqrt{|\Delta|^2+[\mu-U(r) \pm (\hbar^2 v^2_{x}k^2_{x}+\hbar^2 v^2_{y}k^2_{y})^{\frac{1}{2}}]^2}
\label{And-Spe}
\end{eqnarray}

From Eq.(\ref{DBdG}), one can find the wave function in the normal and superconducting region. In the normal region, for electrons (holes) traveling in the $\pm x~~(-x)$ direction with a transverse momentum $k_{y}$ and excitation energy $\epsilon$, the wave functions are given by
\begin{eqnarray}
\Psi^{e+}_{N}&=&(1,e^{i\theta^e_{N}},0,0)^Texp(ik^{e}_{x}x)\nonumber\\
\Psi^{e-}_{N}&=&(1,-e^{-i\theta^e_{N}},0,0)^Texp(-ik^{e}_{x}x)\nonumber\\
\Psi^{h-}_{N}&=&(0,0,1,e^{-i\theta^A_{N}})^Texp(-ik^{h}_{x}x)
\label{wave-function}
\end{eqnarray}
where $k_{x(y)}^{e(h)}=p_{x(y)}^{e(h)}/\hbar$ and we define,
\begin{eqnarray}
e^{i\theta^e_{N}}&=&\frac{v_{x}\cos\theta_{e}+iv_{y}\sin\theta_{e}}{v^e_{0}}\nonumber\\
e^{i\theta^A_{N}}&=&\frac{v_{x}\cos\theta_{h}+iv_{y}\sin\theta_{h}}{v^h_{0}}\nonumber\\
{v_{0}^{e(h)}}&=&\sqrt{v^2_{x}\cos^2\theta_{e(h)}+v^2_{y}\sin^2\theta_{e(h)}}~~.
\end{eqnarray}
Here $\theta_{e}$ is the angle of incidence of the electron and $\theta_{h}$ is the Andreev reflected angle for a hole across the interface. 
Due to the anisotropy in the spectrum ($i.e.,~v_x\ne v_y$), $\theta^e_N\ne\theta_e$, in general and thus the pseudospin-momentum locking
of the unirradiated graphene gets lost here. Rather a modified $\alpha$-dependent relation exists between pseudospin and ${\bf k}$
directions. The critical angle for Andreev reflection $(\theta_{c})$ turns out to be
\begin{eqnarray}
\theta_{c}=\sin^{-1}\frac{\frac{|\epsilon-\mu|}{\epsilon+\mu}v_{x}}{\sqrt{(\frac{|\epsilon-\mu|}{\epsilon+\mu})^2 v^2_{x}+v^2_{y}
(1-(\frac{|\epsilon-\mu|}{\epsilon+\mu})^2 )}}.
\end{eqnarray}
Note that, in a normal graphene (in absence of dressing field) $v_{x}=v_{y}=v_{F}$ and the value of $\theta_{c}$ is given in Ref.\onlinecite{Beenakker-PRL06}.

In the superconducting region, the BdG equation describes the electron and hole quasiparticle mixture or Bogoliubons and opens a gap at the Fermi level. 
There the wave functions take the form
\begin{align}
\Psi^{+}_{S}&=(u(q_{e}),u(q_{e}) e^{i\theta^e_{S}},v(q_{h}),v(q_{h})e^{i\theta^e_{S}})^T exp(iq^{e}_{x}x)\nonumber\\
\Psi^{-}_{S}&=(v(q_{h}),-v(q_{h})e^{-i\theta^h_{S}},u(q_{e}),-u(q_{e})e^{-i\theta^h_{S}})^Texp(-iq^{h}_{x}x)
\end{align}
where
\begin{align}
u(q_{e})&=\sqrt{\frac{1}{2}(1+\frac{\sqrt{\epsilon^2-\Delta^2_{0}}}{\epsilon})},
v(q_{h})=\sqrt{\frac{1}{2}(1-\frac{\sqrt{\epsilon^2-\Delta^2_{0}}}{\epsilon})},\nonumber\\
e^{i\theta^e_{S}}&=\hbar(v_{x}q^e_{x}+iv_{y}k_{y})/(\mu+U_{0}+\Omega),\nonumber\\
e^{-i\theta^h_{S}}&=\hbar(v_{x}q^e_{x}+iv_{y}k_{y})/(\mu+U_{0}-\Omega),~
\Omega=\sqrt{\epsilon^2-\Delta^{2}_{0}}.
\end{align}
Wave-vectors $q^{e}_{x}$ and $q^{h}_{x}$ can be obtained from Eq.(\ref{And-Spe}) in the superconducting region.

Now for an electron incident at the junction from the normal side, and
 with excitation energy $\epsilon$ and transverse momentum $p_{y}$, the wave functions 
in the normal and superconducting regions, taking into account both Andreev and normal reflection processes, can be written as,
\begin{eqnarray}
\Psi_{N}&=&\Psi^{e+}_{N}+r\Psi^{e-}_{N}+r_{A}\Psi^{h-}_{N}\nonumber\\
\Psi_{S}&=&t\Psi^{+}_{S}+t^{'}\Psi^{-}_{S}
\end{eqnarray}
where $r$ and $r_{A}$ are the amplitudes of normal and Andreev reflection respectively, $t$ and $t^{'}$ are the amplitudes of electron-like 
and holelike quasiparticles in the superconducting region. Here, $k^{e(h)}_{x}=k^{N}_{e(h)}\cos\theta$ is the x-component of momentum 
which is not conserved due to broken translational symmetry, whereas $k_{y}=k^{N}\sin\theta$ is conserved. These wave functions must satisfy
the boundary condition,
\begin{eqnarray}
\Psi_{N}(x=0)=\Psi_{S}(x=0)
\end{eqnarray}
Using the boundary conditions one can now solve for the coefficients $r$ and $r_{A}$ to obtain
\begin{eqnarray}
r&=&\frac{u(e^{i\theta^e_{N}}-e^{i\theta^e_{S}})+v \Gamma(e^{i\theta^e_{N}}+e^{-i\theta^h_{S}})}{D},\nonumber\\
r_{A}&=&\frac{2\cos \theta^e_{N}(v+u\Gamma)}{D},\nonumber\\
D&=&u(e^{i\theta^e_{S}}+e^{-i\theta^e_{N}})+v \Gamma(e^{-i\theta^e_{N}}-e^{-i\theta^h_{S}}),\nonumber\\
\Gamma&=&\frac{v(e^{i\theta^e_{S}}-e^{-i\theta^{A}_{N}})}{u(e^{-i\theta^A_{N}}+e^{-i\theta^{h}_{S}})}.
\label{rA-eq}
\end{eqnarray}
The differential conductance of the $NS$ junction follows from the Blonder-Tinkham-Klapwijk formula
\begin{align}
G/G_{0}=\int^{\pi/2}_{0}[1-|r(\epsilon,\theta,\alpha)|^2+|r_{A}(\epsilon,\theta,\alpha)|^2|]\times \cos(\theta) d\theta .
\end{align}
where $G$ is the conductance across the NS junction and $G_0$ is the ballistic conductance of metallic graphene\cite{subhro}.
\begin{figure}[t]
%\vskip .2 in
\center
\rotatebox{0}{\includegraphics[width=3.4in]{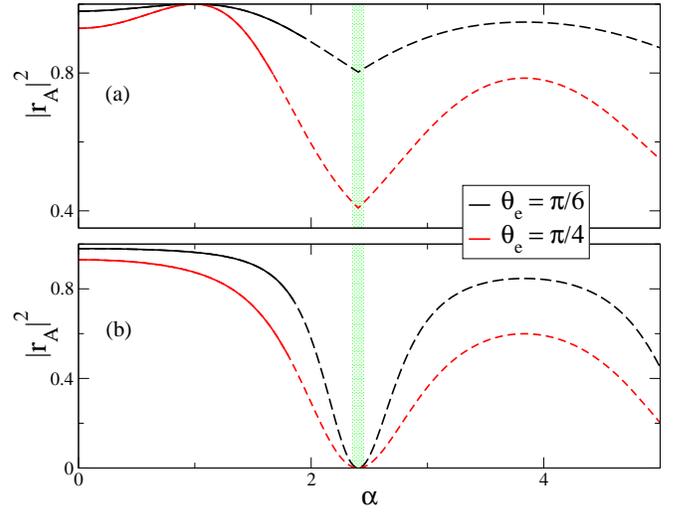}}
\caption{Plot of $|r_A|^2$ with $\alpha$ for different values of $\theta_e$  when the direction of the field is given 
by {{(a)}} $\theta_0=0$ and {{(b)}} $\theta_0=\pi/2$. The parameters are set as 
$\epsilon=0.02\Delta_{0}$, $\mu=100\Delta_{0}$ and $U_0=30\Delta_{0}$. The shaded/dotted zones denote the regions at and around 
$J_0(\alpha)=0$ which are beyond the scope of the present theory.}
\label{reflection-coeff}
\end{figure}
%\begin{figure}[t]
%\vskip .2 in
%\center
%\rotatebox{0}{\includegraphics[width=2.5in]{reflection-coeff.pdf}}
%\caption{Plot of $|r|^2$ with $\alpha$ for different values of $\theta$. The parameters are set as $\epsilon=0.02\Delta_{0}$, $\mu=10\Delta_{0}$ and $U=100\Delta_{0}$. The blue, red and black curves are for $\theta=\pi/6$, $\pi/4$ and $\pi/3$ respectively.}
%\label{reflection-coeff}
%\end{figure}
%\begin{figure}
%\center
%\rotatebox{0}{\includegraphics[width=2.5in]{Andreev-refl.pdf}}
%\caption{Plot of $|r_{A}|^2$ with $\alpha$ for different values of $\theta$. The parameters are set as $\epsilon=0.02\Delta_{0}$, $\mu=10\Delta_{0}$ and $U=100\Delta_{0}$. The blue, red and black curves are for $\theta=\pi/6$, $\pi/4$ and $\pi/3$ respectively.}
%\label{Andreev-refl}
%\end{figure}

% \red{Before going to discuss our results, we mentioned that Andreev reflection in graphene may be retro or specular depending on the value of subgap energy and
% chemical potential ($\mu$). For $\epsilon \leq \Delta_{0}$, the retro-reflection becomes significant if $\mu \gg \Delta_{0}$ and specular reflection dominates 
% if $\mu \ll \Delta_{0}$. Here we consider a regime, $\mu \gg \Delta_{0}$ which is easiest way to reach experimentally.}

{As we see the optical effect to sprout from the 
anisotropy in the prefactors $v_x,~v_y$ (with the ratio being $J_0(\alpha)$), we first 
probe the effect of the dimensionless optical
parameter $\alpha=\frac{2ev_FE_0}{\hbar\omega^2}$ on the andreev/normal reflectance.}
Fig(\ref{reflection-coeff}) shows the variation of probability for Andreev reflection ($|r_{A}|^2$) with $\alpha$ for different values of $\theta_e$. 
In the regime of $\epsilon<\Delta_0$, no quasiparticle transport occurs 
across the NS junction. However, transmission occurs due to Andreev reflection maintaining the constraint $|r|^2+|r_A|^2=1$. 
In Fig.\ref{reflection-coeff}(a) we show the variation of $|r_{A}|^2$ for polarization angle $\theta_0=0$ while
Fig.\ref{reflection-coeff}(b) shows the same for $\theta_0=\pi/2$.
The plots are shown for $\alpha$ upto 5 showing the variation of the Andreev reflectance, even though we put our emphasis only on
small values (shown by solid lines) of $\alpha$ for which our approximation works the most. 
In presence of light, the quasiparticles feel additional force along (or opposite to) the electric field direction and 
accordingly for $\theta_0=0$, they bend towards $x$ direction 
while  for $\theta_0=\pi/2$, they bend towards the $y$ direction.
The reduction in the $v_y$, in the first case, help keeping $r_A$ large due to Klein tunneling while reduction in $v_x$ in the latter case cause $r$ to increase
and $r_A$ to diminish (in fact, it causes $r_A=0$ when $v_x$ vanishes, though this point lies outside the scope of the 
present theory).
\begin{figure}
\center
\rotatebox{0}{\includegraphics[width=3.6in]{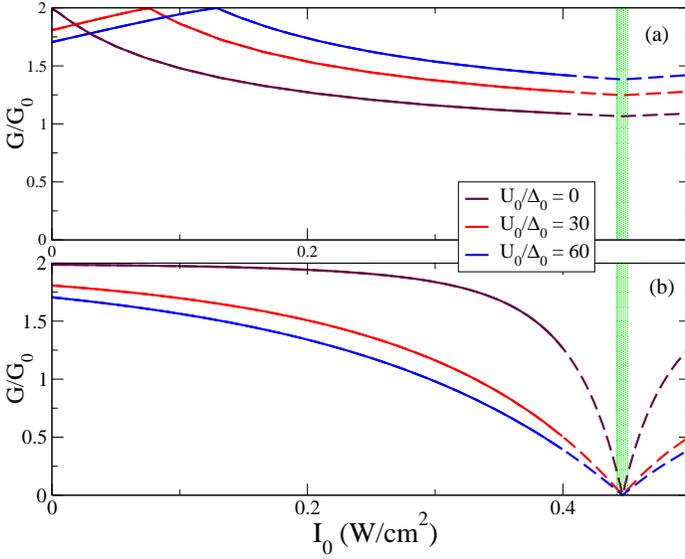}}
\caption{Plot of differential conductance $(G/G_{0})$ of N-S junction versus light intensity $\rm{I}_0$ for
$U_{0}=0, 30$ and $60$ (magenta, purple and pink) respectively with (a) $\theta_0=0$ and (b) $\theta_0=\pi/2$. Here 
we use $\epsilon=0.02~\Delta_{0}$ and $\hbar\omega$=1 meV. The shaded/dotted regime denotes the zone which are beyond the scope of
the present theory. Also the optimally working limit of our theory is the small $I_0$ limit which is shown schematically by the solid lines
(instead of dashed lines that follows).}
\label{Floquet-Conduct}
\end{figure}
In the sub-gap limit, the differential conductance also remains a functional of $r_A$ (or, $r$) alone.  Additionally for $\epsilon<(>)\mu$, retro (specular) type of Andreev reflection develops at the 
junction\cite{ben-rmp}. The condition $\mu>>\Delta_0$ implies $\mu>>\epsilon$ and retro-reflection is obtained. 
Note that, we indeed consider this regime of $\epsilon<\Delta_0$ and $\Delta_0 \ll \mu$ in this paper as that is what is achieved comfortably in experiments.
However, for the sake of continuity of discussion we also point out that for $\mu<<\Delta_0$, $\epsilon>\mu$ is obtained only within the restricted range
of $\Delta_0>\epsilon>\mu$ when specular Andreev reflection is observed\cite{efetov}.
As $\epsilon$ becomes larger than $\Delta_0$, normal tunneling begins and Andreev tunneling becomes smaller and smaller.
%Strictly speaking, in the limit of $\Delta_0>\epsilon>E_F$,
%specular Andreev reflection is observed\cite{efetov}. As $\epsilon$ becomes larger than $\Delta_0$, normal tunelling begins and Andreev tunelling becomes smaller and smaller.
\begin{figure}[b]
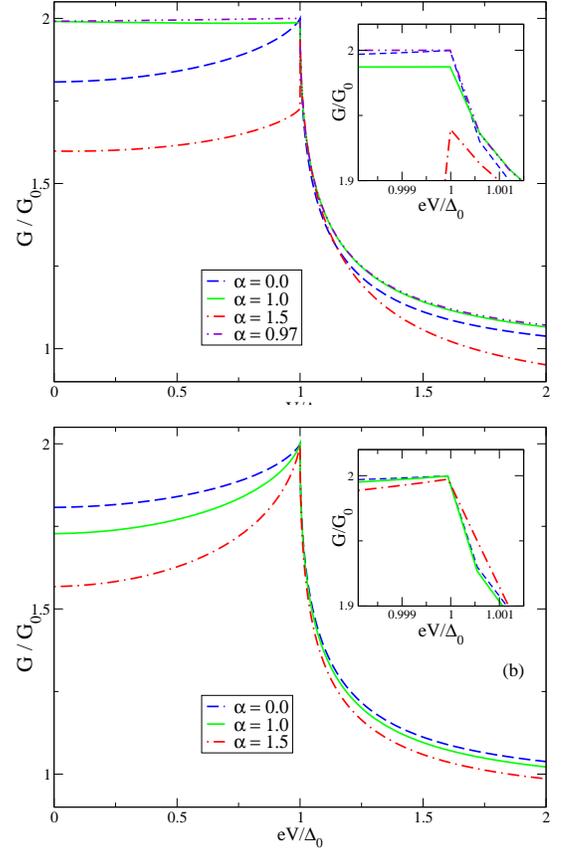

\center
\rotatebox{0}{\includegraphics[width=2.8in]{cond-vs-eV3.eps}}\\
\rotatebox{0}{\includegraphics[width=2.8in]{cond-vs-eV2.eps}}
\caption{Plot of differential conductance $(G/G_{0})$ of N-S junction as a function of $eV/\Delta_{0}$ for 
$\alpha=0$ , 1 and 1.5 (blue, green and red) respectively with
(a) $\theta_0=0$ and (b) $\theta_0=\pi/2$. Here we use $U_{0}=30$ and $\Delta_{0}=0.01\mu$. The inset zooms in
the region at the conductance peak. Furthermore additional plot at $\alpha=0.97$ is shown in (a) where complete subgap
conductance is observed.}
\label{energy-cond}
\end{figure}

Let us first discuss shortly the features of the transport phenomena in normal graphene NS junction (i.e., in absence of linearly 
polarized light or $\alpha=0$). Without a junction, we will neither have electron reflection nor andreev reflection henceforth yielding $G=G_0$. 
Now putting a superconductor-interface there witness andreev transport, for $\epsilon<\Delta_0$, thereby
increasing the conductivity $G$. Without any potential barrier ($i.e.,~U_0=0$),
$G$ maximizes to $2G_0$ when no electron reflection takes place at the junction.
%Andreev reflection causes $G$ to increase with $\epsilon$ as long as $\epsilon\le\Delta_0$, because a larger subgap 
%excitation tends more to get Andreev reflected and pushes more cooper pairs into the S region. For $\mu>>(\epsilon,\Delta_0)$, that we 
%consider, the amplitude $|r_A|$ keeps increasing with $\epsilon$ and becomes unity at $\epsilon=\Delta_0$ where the normal reflection ceases to exist. 
For larger values of $\epsilon$ beyond $\Delta_0$, the electronic system becomes purely resistive and $|r_A|$ gradually decreases
down to zero (see Eq.\ref{rA-eq} for the expressions for $r_A$). 
%Without a fermi surface mismatch ($i.e.,~U_0=0$), incident electrons
%with $\epsilon\le\Delta_0$ faces the gap in the S region and can only cause superconduction via Andreev tunnelling upto a maximum of 
%$G/G_{0}=2$. But a finite barrier develops at the interface when $U\ne0$ and consequently only a fraction of incident electrons 
%Andreev-reflect (and the rest reflect normally) to reduce the conductivity. 
As normal reflection probability increases with barrier height,
a larger $U_0$ results in lesser amount of Andreev reflection and reduced conductivity (see $\alpha=0$ point in Fig.2). For $\epsilon>>\Delta_0$, only a small 
fraction $\sim~\Delta_0/\epsilon$ of the incident electrons get Andreev reflected\cite{bt}.

Now let us take a look at what a tuning via light irradiation can cause to this transport phenomena. 
% \red{Fig(\ref{reflection-coeff}) shows the variation of probability for normal reflection ($|r|^2$) and Andreev reflection ($|r_{A}|^2$) with dimensionless
% optical parameter $\alpha=\frac{2ev_FE_0}{\hbar\omega^2}$ for different values of $\theta$.  In the regime of $\epsilon<\Delta_0$, no quasiparticle transport
% occurs across the NS junction. However, transmission occurs due to Andreev reflection resulting in $|r|^2+|r_A|^2=1$, as seen in Fig.(\ref{reflection-coeff}).}
With moderate $U_0$, the andreev reflection or $r_A$ is generally large for 
$\epsilon<\Delta_0$. But with optical irradiation, andreev transport get reduced in a 
periodic manner, as shown in Fig.1 (for $r_A$) and Fig.2 (for conductivity). However, for $\theta_0\rightarrow 0$ and for small $\alpha$ we see an opposite trend in $r_A$ or $G$. For finite $U_0$, it first increases with $\alpha$ and
start decreasing in an oscillatory fashion only after attaining the maximum
at an intermediate $\alpha$ value (see Fig.1(a) and Fig.2(a)). 
% Such reduction sprouts from the modification of $v_y$ by the factor $J_0(\alpha)$ which makes 
% quasiparticle movement along the $y$ direction slower resulting in lesser amount of conduction.
We should point out here that $|r_A|^2$ smoothly goes towards zero for $J_0(\alpha)\rightarrow0$, 
as $v_y\rightarrow 0$ in those cases allowing mostly the reflection to happen (only for $U_0\ne0$, 
whereas for $U_0=0$, $r$ takes a sharp jump from its minimum to unity where $J_0(\alpha)=0$).
However, this discussion is redundant as our theory breaks down in such limit as many $n'\ne0$ terms from 
the summation in Eq.~\ref{fl-eq} become significant rendering our rotating wave approximation\cite{nori} 
type formalism invalid.
%However, the zeroes of the Bessel's function take us out of the region of validity of the approximation partaken. 
% \red{Though such exact minimum points are inaccessible by our Floquet theory, we can get an impression 
% of how $|r_A|$ or $(G/G_0)$ can be tuned to small values using electromagnetic radiations.} 
As discussed for the $\alpha=0$ case, here also a finite $U_0$ or Fermi level mismatch results in a 
reduction in andreev reflection and conductivity. In Fig.\ref{Floquet-Conduct}, we have taken the frequency of incident 
light to be $\hbar\omega$=1 meV and plotted $G/G_0$ against the light intensity
$\rm{I}_0=\frac{1}{2}cE_0^2$.

In Fig.\ref{energy-cond},
we show the $\epsilon$ dependence of the conductance. With an increase in the 
excitations above the Fermi level,
conductivity increases due to enhanced andreev reflection which becomes maximum at $\epsilon=\Delta_0$.
The $\alpha$ dependence that we saw previously in Fig.\ref{Floquet-Conduct} for very small $\epsilon$, survives
for larger $\epsilon~(<\Delta_0)$ values as well. At $\epsilon=0$, 
conductivity starts from a finite value that is a function of both $\alpha$ 
and $U_0$. As excitation energy increases, so does the Andreev current 
(for $U_0\ne 0$) and $G/G_0$ increases gradually until $\epsilon=\Delta_0$
when the incident quasiparticles no more face any gap at the boundary.
Beyond that point $G/G_0$ show a resistive decay. The critical point
$\epsilon=\Delta_0$ also witness a fine reduction in conductance from its maximum value 2 as $\alpha$ becomes nonzero
(see the inset in Fig.\ref{energy-cond}). For $\theta_0=0$, subgap
conductivity remains maximum for one optimum value of $\alpha$
which varies with $U_0$ (as seen in Fig.\ref{Floquet-Conduct}(a)). In fact, the corresponding upturn in conductivity 
as $\alpha$ is turned on gradually, is responsible for the sudden jump in $G/G_0$ at/near $\epsilon=\Delta_0$ as seen 
in Fig.\ref{energy-cond}(a). Mean field condition for 
superconductivity\cite{Beenakker-PRL06}, $i.e.~~\mu+U_0>>\Delta_0$ is considered throughout all calculations 
to ensure that phase coherence is maintained within the S region over a
distance of $\lambda_F^S=\frac{\hbar v_F}{\mu+U_0}$. 

%{We should point out here that we obtain $|r|(|r_A|)\rightarrow 1(0)$ for $J_0(\alpha)\rightarrow0$ as $v_x\rightarrow 0$ in those cases
%allowing mostly the reflection to happen (only for $U_0\ne0$. For $U_0=0$, however, $r$ takes a sharp jump from 0 to 1 where $J_0(\alpha)=0$).
%However, this discussion is redundant as our theory breaks down in such limit as it becomes impossible to single out one term from 
%the summation in Eq.~\ref{fl-eq} and thus rotating wave approximation\cite{nori} like formalism fails in obtaining Eq.~\ref{fl-eq2} from there.}

In the next section, we will discuss about the ABS and Josephson current in an irradiated graphene SNS junction.

\section{Andreev bound states and Josephson current}
\begin{figure}
\vskip .2 in
\includegraphics[height=1.5 in, width=3.4 in]{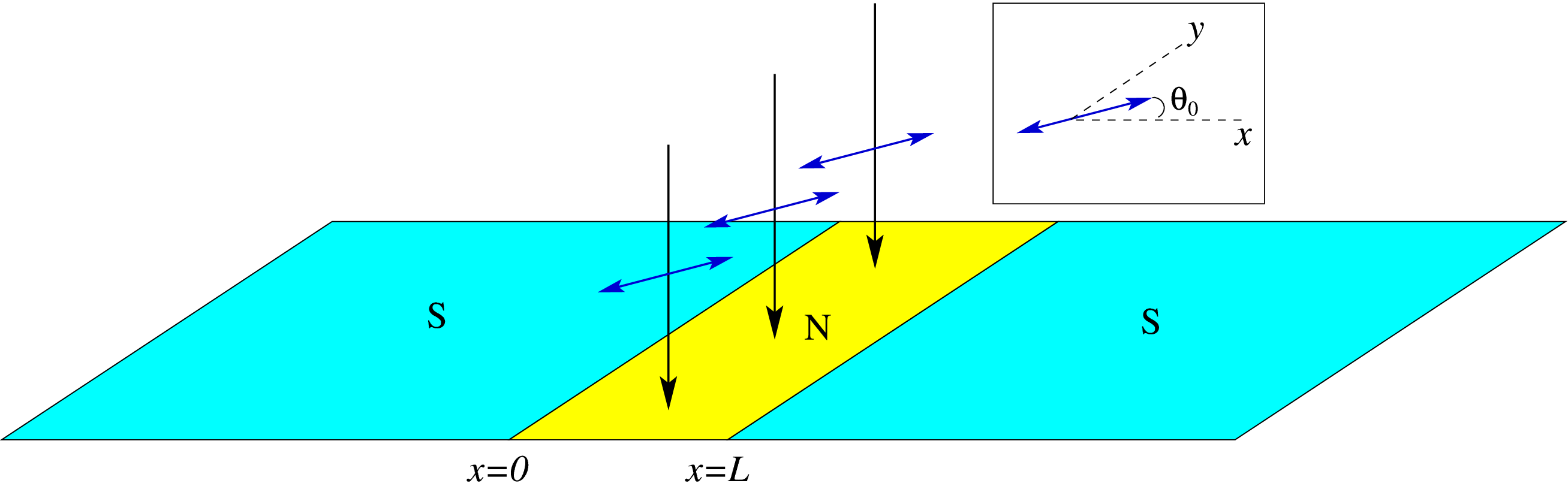}
\caption{Light irradiation to a SNS junction with electric field at an angle $\theta_0$ to the $x$ axis (the junction direction 
in the $xy$ plane).}
\label{sns}
\vskip -.2 in
\end{figure}

Josephson supercurrent develops in a SNS junction due to proximity effect\cite{proxi}
and this is expressed in terms of the quantized andreev bound states (ABS)
developed within the intermediate normal region.
To calculate ABS and Josephson current of optically dressed electrons, we consider a irradiated graphene SNS junction where superconducting electrodes 
are
deposited in the left (region-I, with $x<0$) and right regions (region-II, with $x>L$), leaving a narrow middle region (II) to be the normal 
graphene (see Fig.\ref{sns}).
{What we describe below is a brief narrative of graphene SNS junction calculations of Ref.\onlinecite{Beenakker-PRL06,
subhro}, worked out for our present case incorporating velocity anisotropy.}
Under exposure to linearly polarized light, the energetics in the three different regions get modified and we construct the
wave functions as follows. For $x<0$, we may have,
\begin{align}
\Psi_{S,L}&=t^e_{L}e^{-iq_{e}\cos\theta^{e}_{S}x}
[u(q_{e}),
u(q_{e})e^{i(\pi-\theta^{e}_{S})},
v(q_{e})e^{-i\phi_{L}},\nonumber\\
&v(q_{e})e^{i(\pi-\theta^{e}_{S}-\phi_{L})}]^T+t^h_{L}e^{iq_{h}\cos\theta^{h}_{S}x}
[v(q_{h}),\nonumber\\
&v(q_{h})e^{i\theta^{h}_{S}},
u(q_{h})e^{-i\phi_{L}},
u(q_{h})e^{i(\theta^{h}_{S}-\phi_{L})}]^T
\label{wave-L}
\end{align}
and for $x>L$,
\begin{align}
\Psi_{S,R}&=t^e_{R}e^{iq_{e}\cos\theta^{e}_{S}x}
[u(q_{e}),u(q_{e})e^{i\theta^e_{S}},v(q_{e})e^{-i\phi_{R}},\nonumber\\
&v(q_{e})e^{i(\theta^e_{S}-\phi_{R})}]^T
+t^h_{R}e^{-iq_{h}\cos\theta^{h}_{S}x}
[v(q_{h}),\nonumber\\&v(q_{h})e^{i(\pi-\theta^{h}_{S})},u(q_{h})e^{-i\phi_{R}},u(q_{h})e^{i(\pi-\theta^{h}_{S}-\phi_{L})}]^T.
\label{wave-R}
\end{align}
In the region $0\le x \le L$,  we can construct wave function as given in Eq.(\ref{wave-function}). Here $\phi_{L,R}$ is the superconducting 
phase on the left/right side of the normal region, associated with the broken $U(1)$ symmetry in the superconducting state. The macroscopic 
phase difference is defined as $\phi=\phi_{R}-\phi_{L}$. The procedure for calculating the Josephson current is to first obtain
the energy spectrum for the Andreev bound states in the intermediate normal region. This is done by matching the wavefunctions at the two NS 
interfaces, and then solving for the allowed energy states. Explicitly, the boundary conditions dictate that,
\begin{eqnarray}
\Psi^{S}_{L}(x=0)=\Psi^{N}(x=0); \Psi^{S}_{R}(x=L)=\Psi^{N}(x=L)
\label{boundary-conds}
\end{eqnarray} 
leading to quantization relations between the superconducting phase difference $\phi$ and the quasiparticle excitation energy $\epsilon$.
The boundary conditions in Eq.(\ref{boundary-conds}) lead to a matrix equation involving an $8\times 8$ matrix\cite{lind2}
\begin{eqnarray}
\mathcal{M}=\begin{pmatrix}
\mathcal{M}_{11} & \mathcal{M}_{12}\\
\mathcal{M}_{21} & \mathcal{M}_{22}
\end{pmatrix}
\label{Matrix-Form}
\end{eqnarray}
for transmission and reflection coefficients and the non-trivial solutions exist for $det(\mathcal{M})=0$.
\begin{figure}[t]
\center
\rotatebox{0}{\includegraphics[width=2.6in]{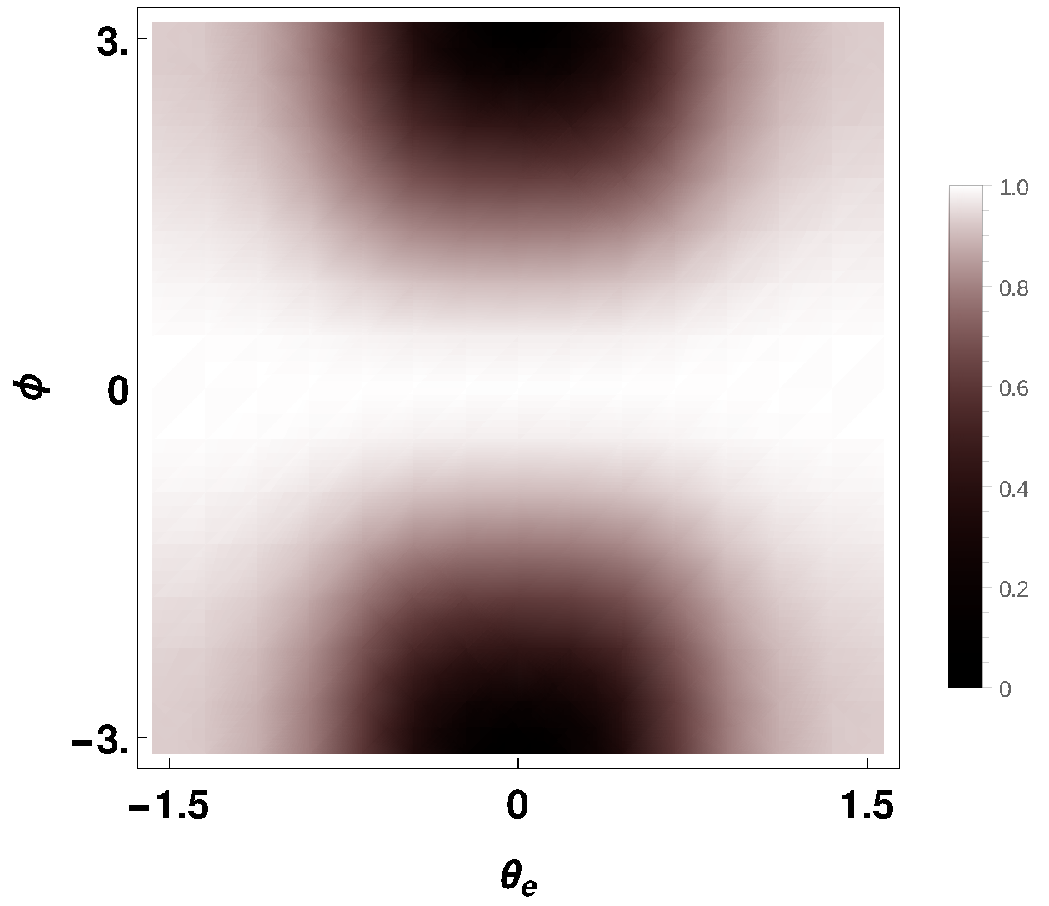}}\\
\vskip .1 in
\rotatebox{0}{\includegraphics[width=1.68in]{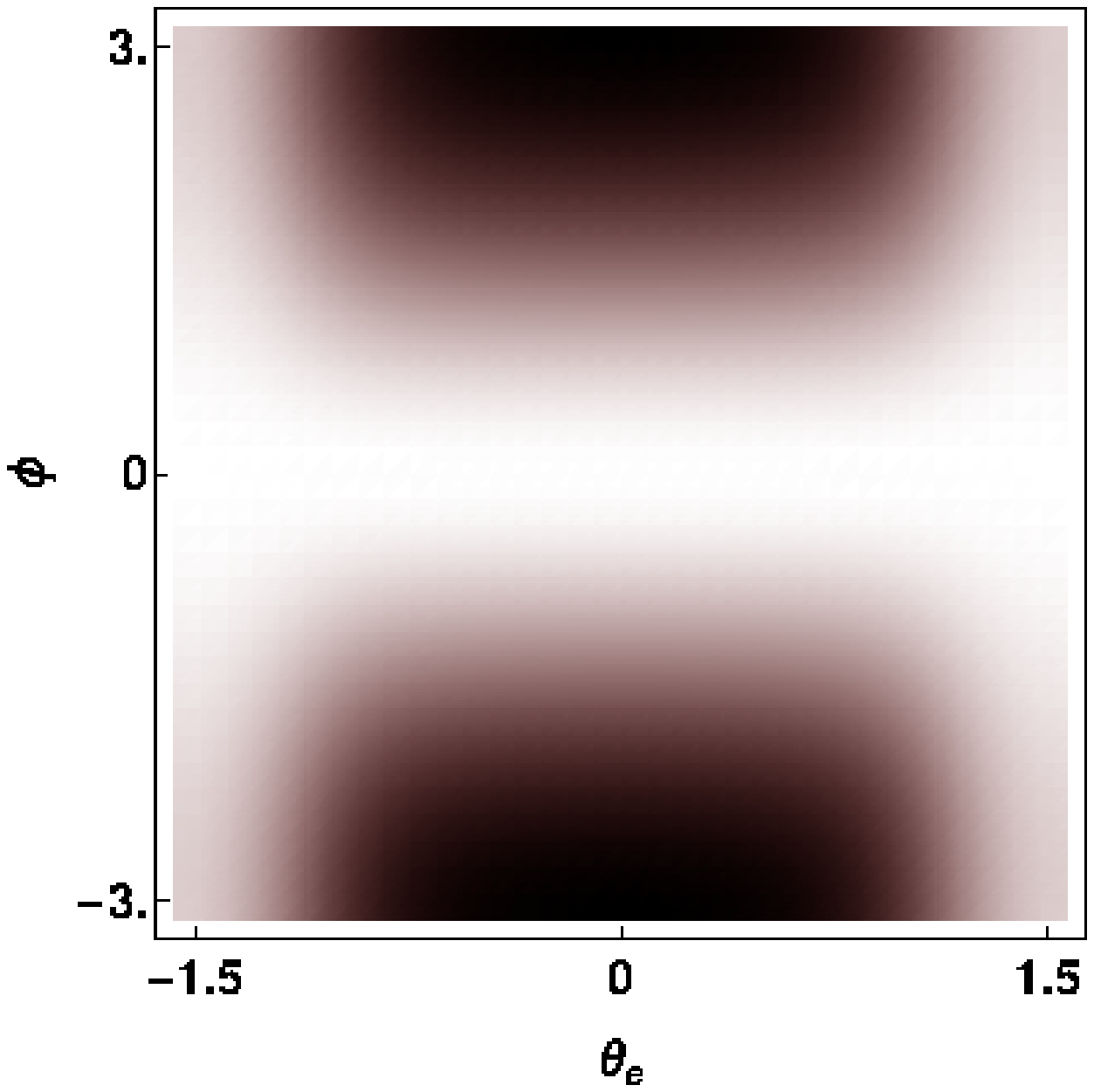}}
\rotatebox{0}{\includegraphics[width=1.68in]{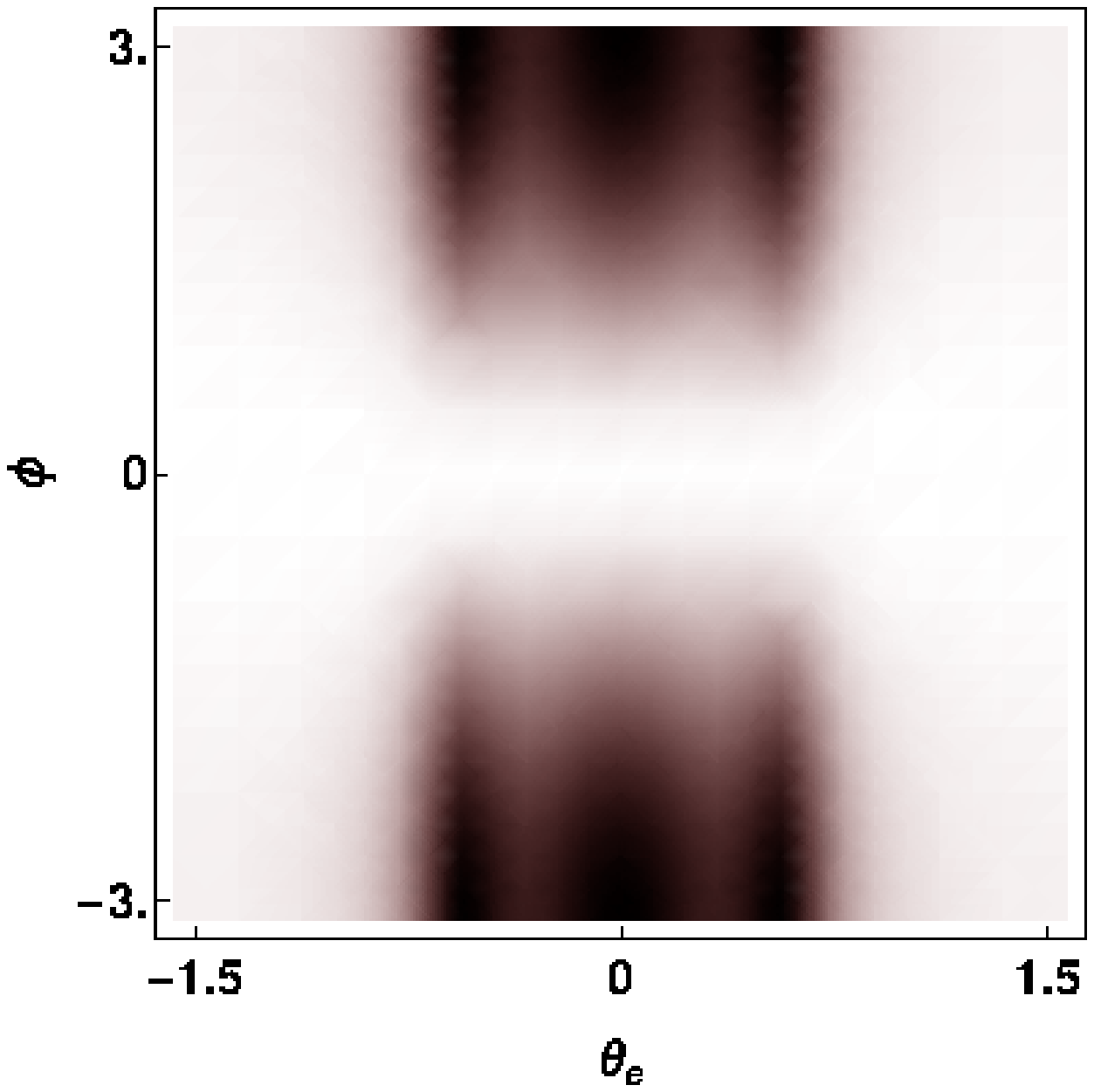}}
\caption{Plot of ABS in the Josephson S$\vert$N$\vert$S junction with $\phi$ and $\theta_e$ in (top) absence and 
(bottom) presence of a dressing field with $\alpha=1.5$. We have set $\mu L/\hbar v_F=2.5$. The field directions
are considered to be along $x$ (bottom-left) and $y$ (bottom-right) respectively.}
\label{Andreev-Bound}
\end{figure}
The $4\times 4$ matrices $\mathcal{M}_{ij}$ are explicitly given in appendix. In order to proceed with the analytical solution we assume the
superconducting regions to be heavily doped, $i.e.$, $\mu_{s} \gg \mu$ (here $\mu_s=\mu+U_0$,
 is the effective chemical potential within the S regions). In this case, 
$\theta^{e}_{S}=\theta^{h}_{S}=\delta$ and the number of propagating modes in superconducting region becomes $N=\mu_{s}W/\pi \hbar v_F$
where $W$ denotes the width of the N region with $W \gg L$. In the regime of 
"short-junction" limit ($\Delta_0L/\hbar v_F \ll 1$) and heavily doped superconductor, one can effectively put $\delta \rightarrow 0$. Within
the approximations, the quantized states are obtained in terms of $\theta$ and $\mu$ of the N region. The expression is obtained as, 
\begin{eqnarray}
\epsilon_{n}(\phi)=\Delta_{0} \sqrt{1-\gamma_{n}\sin^2\phi/2}
\end{eqnarray}
where $\gamma_{n}$ is the transmission probability through the middle region and is obtained as,
\begin{eqnarray}
\gamma_{n}&=&\frac{k^2_{x}}{k^2_{x}\cos^2(k_{x}L)+\frac{\mu^2}{\hbar^2v^2_F}\sin^2(k_{x}L)}\nonumber\\
k_{x}L&=&\sqrt{(\frac{\mu L}{\hbar v_F})^2-J^2_0(\alpha)q_{n}^2L^2}
\label{gamma}
\end{eqnarray}
Here $k_x$ and $q_n$ denote the wave-vectors along $x$ and $y$ directions respectively where the transverse wavevectors, for 
infinite mass confinement\cite{ben2}, are quantized as 
$q_n=(n+\frac{1}{2})\pi/W$. The resonant electron-hole states represented by these
$\epsilon_n(\phi)$ are called the andreev bound states (for $|\epsilon_n(\phi)|<\Delta_0$). Note that for $\alpha=0$, 
Eq.(\ref{gamma}) is reduced to the form as given in Ref.\onlinecite{titov}. The Andreev 
modes collectively contribute to the Josephson supercurrent as\cite{titov},
\begin{eqnarray}
I(\phi)=\frac{e\Delta_{0}}{\hbar}\sum^{N}_{0}\gamma_{n}\sin \phi /\epsilon_{n}(\phi)
\label{supercurrent}
\end{eqnarray}
In short junction limit, we may replace the summation in Eq.(\ref{supercurrent}) over the quantized modes with an integration: 
$\sum_{n}\rightarrow \frac{W}{2\pi} \int dq_{n}$. 

Now we observe how ABS depends on the optical parameter $\alpha$. Fig.\ref{Andreev-Bound} shows the comparison between $\alpha=0$ 
and $\alpha=1.5$. Without irradiation, electrons incident normally ($i.e.,~\theta_e=0$) causes the bound-state energy to become
zero at $\phi=(-\pi, \pi)$, whereas $\epsilon_n(\phi)=\Delta_0$ ($i.e.,$ andreev modes remains bound no more) when $\theta_e
\rightarrow \pm \pi/2$. With the dressing field, such angular dependence of zero-ABS regime - which contributes most to the
superconduction, spread more for $\theta_0=0$ or become an oscillating pattern for $\theta_0=\pi/2$ (where further null values
are obtained at discrete oblique incident angles when $\gamma_n\rightarrow1$).
These can be seen in Fig.\ref{Andreev-Bound} bottom panels. 
\begin{figure}
\vskip -.2 in
\center
\rotatebox{0}{\includegraphics[width=3.5in]{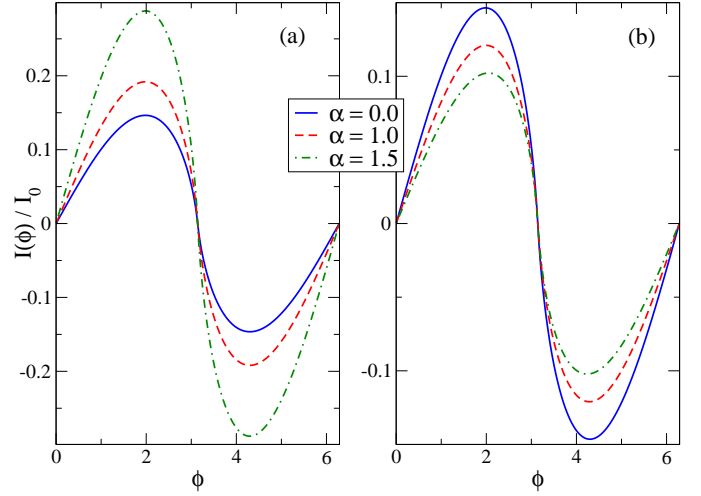}}
%\rotatebox{0}{\includegraphics[width=1.6in]{joseph-y-phi.pdf}}
\caption{Plot of Josephson current $I(\phi)/I_{0}$ as a function of $\phi$ with $\alpha$ as a parameter corresponding to
the electric field of light along (a) $x$-direction and (b) $y$-direction respectively. In both 
the plots we set $\mu L/\hbar v_F=1$, $W/\xi=30$, $\mu_{s}/\Delta_{0}=150$ and $L/\xi=0.1$.}
\label{Joseph-Phi}
\end{figure}

Experimentally, ABS features can also be perceived by noticing the current phase 
relation which is shown in Fig.\ref{Joseph-Phi}.
\begin{figure}
\center
\rotatebox{0}{\includegraphics[width=3.2in]{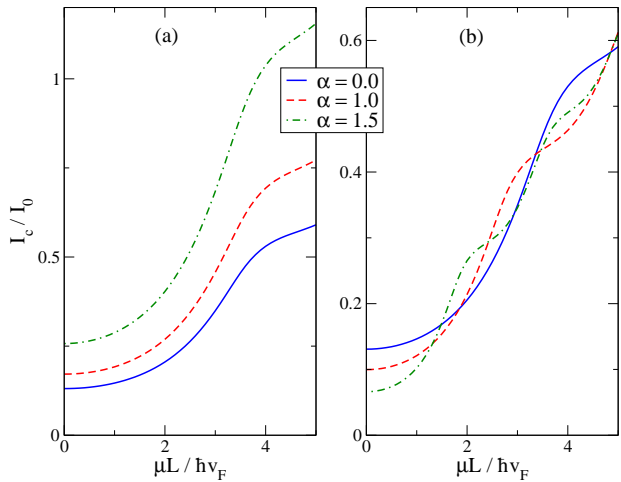}}
\caption{Plot of critical current ($I_{c}/I_{0}$) with $\frac{\mu L}{\hbar v_F}$ with $\alpha$ as a parameter for (a)
$\theta_0=0$ and (b) $\theta_0=\pi/2$. Both 
the plots correspond to $W/\xi=30$, $\mu_{s}/\Delta_{0}=150$ and $L/\xi=0.1$.}
\label{Ic-vs-muL}
\end{figure}
The Josephson current is sensitive to the direction of light irradiation on the graphene. 
This is due to fact that the transmission probability $\gamma_{n}$ is enhanced or suppressed when light radiation is along 
$x$ or $y$-direction respectively.

The behavior of 
critical current $I_{c}$ of irradiated graphene-based Josephson junction vs $\mu L/\hbar v_F$ is shown in Fig.\ref{Ic-vs-muL}. 
Fig.\ref{Ic-vs-muL}(a) shows the plot of $I_{c}/I_{0}$ ($I_0=e\Delta_0/\hbar$) for the light irradiation is along x-direction. As shown, the value of
$I_{c}$ of irradiated graphene becomes larger with finite $\alpha$ as compared to its normal value (for $\alpha=0$) at the 
charge neutrality point. {This is an interesting observation given the fact that graphene transport is already special due to nonzero Josephson current at $\mu=0$ notwithstanding its zero DOS, first shown
analytically by Titov and Beenakker (Ref.\onlinecite{titov}).} When 
light is irradiated along y-direction the expression of $\gamma_{n}$ and $k_{x}$ in Eq.\ref{gamma} is modified as,
\begin{eqnarray}
\gamma_{n}&=&\frac{J^2_{0}(\alpha) k^2_{x}}{J^2_0(\alpha) k^2_{x}\cos^2(k_{x}L)+\frac{\mu^2}{\hbar^2v^2_F}\sin^2(k_{x}L)}\nonumber\\
k_{x}L&=&\frac{\sqrt{(\frac{\mu L}{\hbar v_F})^2-q_{n}^2L^2}}{J_{0}(\alpha)}
\label{gamma-y}
\end{eqnarray}
The value of $I_{c}$ with $\mu L/\hbar v_F$ is shown in Fig.\ref{Ic-vs-muL}(b). It shows that, $I_{c}$ decreases with $\alpha$ at
$\mu \rightarrow 0$. For nonzero $\mu$, however, the value of $I_{c}$ 
and hence Josephson current of irradiated graphene can be either enhanced or suppressed depending on the values of $\mu$ and
$\alpha$ used. All these behavior can be understood examining the expression for $\gamma_{n}$ in Eq.(\ref{gamma}) and Eq.(\ref{gamma-y}). 
For $\mu=0$, all the transport modes $n$ become evanescent and $\gamma_{n}$ decays slower and faster as a function of $q_{n}$ 
in Eq.(\ref{gamma}) and Eq.(\ref{gamma-y}) respectively. These results, to some extent, resemble that from a strained monolayer 
graphene, as detailed in Ref.[\onlinecite{lind2, Wang}]. There, an applied mechanical strain on a monolayer graphene breaks the isotropy 
in the velocity of a quasiparticle by 
modifying the hopping parameter and thereby makes the low energy spectrum anisotropic. This route, therefore, leads to similar outcome as ours.
However, since a light-controlled electronic device is typically much faster and easily controllable than 
those mechanically or electrically controlled devices, an optical tuning stands out as a more feasible means to the experimentalists and
we believe that our findings can encourage a lot during building of future spintronic devices.

% {In a graphene based SNS junction, Josephson current develops via Andreev bound states in the intermediate normal region. 
% Coherent phase dfference
%  $\phi=(\phi_R-\phi_L)$ between the superconducting regions in the left and right side allows for a supercurrent $I(\phi)$ that varies periodically with $\phi$
% with contribution from all possible Andreev modes of closed currrent loops.
% Particularly, in the short junction limit, in which length of the intermediate normal region $L<<\xi$,
% the discrete energy spectrum turns continuous.}
% 
% {We should mention here that all our}
% results are relevant under 
% the condition $|J_n(\alpha)p_ya_n^\pm/J_0(\alpha) p_ya_0^\pm|<<1$ i.e., if the Bessel function $J_{0}(\alpha)$ is far from zero. So, the
% values of $|r|^2$ close to one (in Fig.(\ref{reflection-coeff})) or $|r_{A}|^2$ close to zero (in Fig.(\ref{Andreev-refl})) are not
% applicable physically under our theory. Fig(\ref{Andreev-cond}) shows variation of diffrential conductance of $NS$ junction with dressing
% field. The main feature of the conductance is oscillating behaviour even for large fermi surface mismatch.

\vskip 1 in

\section{Conclusion}
In summary, we have studied extensively the Andreev transport and Josephson effect in light irradiated, proximity induced graphene
NS and narrow SNS junctions. 
In the off-resonant condition that we study, the resulting Dirac spectrum becomes anisotropic 
which unravels many unusual phenomena such as reduction/enhancement of the 
subgap conductance as well as the Josephson currents, depending on the direction of polarization.
In one extreme, we can tune in maximum andreev conductivity in presence of Fermi level mismatch, for irradiated field along 
the junction direction.
On the other hand, noticeable reduction in AR is possible when the polarization
points parallel to the interface/junction.
This latter feature can be utilized in getting enhanced crossed Andreev reflection\cite{guy} (CAR)
in an irradiated graphene based NSN junction. Usually in graphene NSN junctions, CAR is not perceived much
due to the local AR and elastic co-tunneling (EC) processes, unless raising and lowering of the chemical potentials
are performed considering nSp or pSn type graphene bipolar transistors\cite{cayssol}. However, in an irradiated graphene
sheet, AR can be tuned to get considerably suppressed. 
So it will be interesting to investigate the optical effect in strengthening the CAR signal even without shifting the chemical 
potential of the normal leads and thereby causing the non-local cooper pair splitting that
spatially separates the entangled electron pairs\cite{ang}. In fact, such investigation will comprise our future plan of work.
Lastly, irradiation causes redistribution of the low energy regime in the Andreev bound state.
And as we have seen, it can be tuned properly to produce enhanced supercurrent through a graphene based SNS junction,
even at the Dirac point in the spectrum.

These interesting observations, in fact, can provide possible route to control quantum transport in graphene with relevance
to the spintronic based applications.
It will be equally interesting to see the effect of light irradiation in transition metal dichalcogenides such as silicene\cite{jin} 
or $MoS_2$ \cite{neal} where spin orbit interaction plays an important role in transport.
 
% As a matter of fact, optical tunability of the transport behavior has huge advantage over the other conventional tuning mechanisms
% such as straining or creating gaps artificially, due to the ease of experimental implementation in light irradiation.

\section{Acknowledgments}
The authors thank K. Sengupta for useful discussions. SK acknowledges financial support from CSIR, India, under 
Scientists' Pool Scheme No. 13(8764-A)/2015-Pool.
\section{Appendix}
Here, we give the matrix form of $\mathcal{M}_{ij}$ in Eq.(\ref{Matrix-Form}) of the main text. The matrix form can be easily constructed
by matching the wave function in two NS regions and ABS is obtained from the nontrivial solution of the eigenvalue equation $\mathcal{M}x=0$. 
Using Eq.(\ref{wave-function}), Eq.(\ref{wave-L}) and Eq.(\ref{wave-R}), the matrix forms are obtained as,
\begin{widetext}
\begin{align}
\mathcal{M}_{11}&=\begin{pmatrix}
u(q_{e}) & v(q_{h}) & -1 & -1\\
u(q_{e})e^{i(\pi-\theta_{e})} & v(q_{h})e^{i\theta_{h}} & -e^{i\theta} & e^{-i\theta}\\
v(q_{e})e^{-i\phi_{L}} & u(q_{h})e^{-i\phi_{L}} & 0 & 0\\
v(q_{e})e^{i(\pi-\theta_{e}-\phi_{L})} & u(q_{h})e^{i(\theta_{h}-\phi_{L})} & 0 & 0 
\end{pmatrix}
~~~~~~\mathcal{M}_{12}=\begin{pmatrix}
0 & 0 & 0 & 0\\
0 & 0 & 0 & 0\\
-1 & -1 & 0 & 0\\
e^{i\theta_{A}} & -e^{-i\theta_{A}} & 0 & 0
\end{pmatrix}\nonumber\\
\mathcal{M}_{21}&=\begin{pmatrix}
0 & 0 & -e^{ik_{e}\cos\theta x} & -e^{-ik_{e}\cos\theta x}\\
0 & 0 & -e^{i\theta}e^{ik_{e}\cos\theta x} & e^{-i\theta}e^{-ik_{e}\cos\theta x}\\
0 & 0 & 0 & 0\\
0 & 0 & 0 & 0
\end{pmatrix}\nonumber\\
\mathcal{M}_{22}&=\begin{pmatrix}
0 & 0 & u(q_{e})e^{iq_{e}\cos\theta_{e}L} & v(q_{h})e^{-iq_{h}\cos\theta_{h}L}\\
0 & 0 & u(q_{e})e^{iq_{e}\cos\theta_{e}L} e^{i\theta_{e}} & v(q_{h})e^{i(\pi-\theta_{h})}e^{-iq_{h}\cos\theta_{h}L}\\
-e^{ik_{h}\cos\theta_{A}x} & -e^{-ik_{h}\cos\theta_{A}x} & v(q_{e})e^{-i\phi_{R}} e^{iq_{e}\cos\theta_{e}L}&  u(q_{h})e^{-iq_{h}\cos\theta_{h}L}e^{-i\phi_{R}}\\
e^{i\theta_{A}}e^{ik_{h}\cos\theta_{A}x} & -e^{-i\theta_{A}}e^{-ik_{h}\cos\theta_{A}x} & v(q_{e})e^{-i(\theta_{e}-\phi_{R})} e^{iq_{e}\cos\theta_{e}L} & u(q_{h})e^{-iq_{h}\cos\theta_{h}L}e^{i(\pi-\theta_{h}-\phi_{R})}
\end{pmatrix}
\end{align}
\end{widetext}


\begin{thebibliography}{0}
\bibitem{Beenakker-PRL06} C .W. J. Beenakker, Phys. Rev. Lett. 97, 067007 (2006).
\bibitem{ben-rmp} C. W. J. Beenakker, Rev. Mod. Phys. 80, 1337 (2008).
\bibitem{neto} A. H. C. Neto $et~al.$, Rev. Mod. Phys. 81, 109 (2009).
\bibitem{sdsharma} S. D. Sharma $et~al.$, Rev. Mod. Phys. 83, 407 (2011).
\bibitem{andre}K. S. Novoselov, A. K. Geim, S. V. Morozov, D. Jiang, Y. Zhang, S. V. Dubonos, I. V. Grigorieva, A. A. Firsov, Science 306 (5696),pp. 666–669 (2004).
\bibitem{bastos} R. C-Bastos, C. Leon, D. Faria, A. Latge, E. Y. Andrei, N. Sandler, Phys. Rev. B 94, 125422 (2016).
\bibitem{lind2} M. Alidoust, J. Linder, Phys. Rev. B 84, 035407 (2011).
\bibitem{majidi} L. Majidi, M. Zareyan, Phys. Rev. B 86, 075443 (2012).
\bibitem{farhat} M. Farhat, S. Guenneau, H. Bagci, Phys. Rev. Lett. 111, 237404 (2013).
\bibitem{plasmon} J. Schiefele, J. Pedros, F. Sols, F. Calle, F. Guinea, Phys. Rev. Lett. 111, 237405 (2013).
\bibitem{plasmon2} S. Xiao, X. Zhu, B.-H. Li, N. A. Mortensen, Front. Phys. 11(2), 117801 (2016).
\bibitem{iorsh} I. V. Iorsh, I. S. Mukhin, I. V. Shadrivov, P. A. Belov, Y. S. Kivshar Phys. Rev. B 87, 075416 (2013).
\bibitem{kuzmin} O. V. Kibis, S. Morina, K. Dini and I. A. Shelykh, A.Phys.Pol.A 127, 528(2016).
\bibitem{furio} A. Furio $et~al.$, Nanotechnology, 28, 054003 (2017).
\bibitem{Shelykh-1} O. V. Kibis, S. Morina, K. Dini and I. A. Shelykh, Phys. Rev. B, 93, 115420 (2016)
\bibitem{Shelykh-2} K. Kristinsson, O. V. Kibis and I. A. Shelykh, Sci. Report, 6, 20082 (2016);
\bibitem{Shelykh-3} D. Yudin, O. V. Kibis and I. A. Shelykh, NJP 18, 103014 (2016)
\bibitem{Shelykh-4} O. V. Kibis, K. Dini, I. V. Iorsh and I. A. Shelykh, Phys. Rev. B, 95 125401 (2017)
\bibitem{Shelykh-5} D. Yudin, I. A. Shelykh, Phys. Rev. B 94, 161404 (R) (2016)
\bibitem{nori} S. Ashhab $et~al.$, Phys. Rev. A75, 063414 (2007).
\bibitem{zhou} X. Zhou, G. Jin, Phys. Rev. B 94, 165436 (2016).
\bibitem{Luis-1} H. L. Calvo, H. M. Pastawski, S. Roche and L. E. F. Torres, Applied. Phys. Lett 98, 232103 (2011)
\bibitem{Luis-2} G. Usaj, P. M. Perez-Piskunow, L. E. F. Torres and C. A. Balseiro, Phys. Rev. B 90, 115423 (2014)
\bibitem{Jin-2} X. Zhou, Y. Xu and G. Jin, Phys. Rev. B 92, 235436 (2015)
\bibitem{Kibis-1} O. V. Kibis, Phys. Rev. B 86, 155108 (2012)
\bibitem{Kibis-2} S. Morina, O. V. Kibis, A. A. Pervishko, I. A. Shelykh, Phys. Rev. B 91, 155312 (2015)
\bibitem{Kibis-3} A. A. Pervishko, O. V. Kibis, S. Morina, I. A. Shelykh, Phys. Rev. B 92, 205403 (2015)
\bibitem{Kibis-4} O. V. Kibis, Phys. Rev. Lett. 107, 106802 (2011)
\bibitem{Rubio} H. H\"{u}bener  $et~al.$, Nature Communication 8, 13940 (2017)
\bibitem{sinha} D. Sinha, Eur. Phys. Lett. 115, no. 3, 37003 (2016).
\bibitem{Jin-1} X. Zhai and G. Jin, Phys. Rev. B 89, 235416 (2014).
%\bibitem{dahal2} H. P. Dahal, Z.-X. Hu, N. A. Sinitsyn, K. yang, A. V. Balatsky, Phys. Rev. B 81, 155406 (2010)
%\bibitem{neilson} D. Neilson, A. Perali and M. Zarenia, J. Phys. Conf. Ser.702 (2016) 012008.
%\bibitem{cohen} C. Cohen-Tannoudji, J. Dupont-Roc, and G. Grynberg, ``Atom-Photon Interactions: Basic Processes and Applications''(Wiley,Chichester, 1998).
\bibitem{Ezawa} M. Ezawa, Phys. Rev. Lett 100, 026603 (2013)
\bibitem{oka} T. Kitagawa, T. Oka, A. Brataas, L. Fu, and E. Demler, Phys. Rev. B 84, 235108 (2011)


\bibitem{proxi} Heersche, H. B., P. Jarillo-Herrero, J. B. Oostinga, L. M. K.
Vandersypen, and A. Morpurgo, 2007, Nature 446, 56.
\bibitem{linder} J. Linder, A. Sudbo, Phys. Rev. B 77, 064507 (2008).
\bibitem{titov} M. Titov, C. W. J. Beenakker, Phys. Rev. B74, 041402 (2006).
\bibitem{maiti} M. Maiti, K. Sengupta, Phys. Rev. B76, 054513 (2007).
\bibitem{jacob} J. Linder $et~al.$ Phys. Rev. B 80, 094522 (2009)
\bibitem{nori2} S. V. Syzranov, Ya. I. Rodionov, K. I. Kugel and F. Nori, Phys. Rev. B 88, 241112 (2013)
%\bibitem{kris} A. Polkovnikov $et~al.$, Rev. Mod. Phys. 83, 863 (2011).
\bibitem{klein} M. I. Katsnelson, K. S. Novoselov, and A. K. Geim, Nat. Phys. 384, 620 (2006).
\bibitem{subhro} S. Bhattacharjee, K. Sengupta, Phys. Rev. Lett. 97, 217001 (2006).
%\bibitem{efetov} D. K. Efetov $et~al.$, Nat. Phys. 12, 328 (2016).
\bibitem{bt} G. E. Blonder, M. Tinkham, Phys. Rev. B27, 112 (1983).
%\bibitem{Alidoust} M. Alidoust, J. Linder, Phys. Rev. B 84, 035407 (2011)
\bibitem{ben2} J. Tworzydlo $et~al.$, Phys. Rev. Lett. 96, 246802 (2006).
\bibitem{Wang} Y. Wang, Y. Liu and B. Wang, Appl. Phys. Lett. 103, 182603 (2013)
\bibitem{guy} G. Deutscher, Jour. of Sup., vol 15, no.1, page 43 (2002)
\bibitem{cayssol} J. Csyssol, Phys. Rev. Lett. 100, 147001 (2008).
\bibitem{ang} Y. S. Ang, L. K. Ang, C. Zhang, and Z. Ma, Phys. Rev. B93, 041422(R) (2016)
\bibitem{jin} X. Zhou, and G. Jin, Phys. Rev. B94, 165436 (2016).
\bibitem{neal} A. T. Neal, H. Liu, J. Gu, and P. D. Ye, ACS Nano 7(8), pp 7077 (2013).
\end{thebibliography}
\end{document}